# Single-pass generation of widely-tunable frequency-domain entangled photon pairs

**MASAYUKI HOJO** [1, *] **AND KOICHIRO TANAKA**[1, 2]

[1] *Department of Physics, Science, Kyoto University, Kitashirakawa-Oiwake, Sakyo, Kyoto, Japan*
[2] *Institute for Integrated Cell-Material Sciences (iCeMS), Kyoto University, Sakyo, Kyoto, Japan*
*hojo.masayuki.99n@kyoto-u.jp

**Abstract:** We demonstrate a technique that generates frequency-entangled photon pairs with high polarization definition by using a single-period nonlinear crystal and single pass configuration. The technique is based on the simultaneous occurrence of two spontaneous parametric down-conversion processes satisfying independent type-II collinear pseudo-phase matching conditions in periodically poled stoichiometric lithium tantalate. The generated photon pairs exhibit non-degenerate Hong-Ou-Mandel interference, indicating the presence of quantum entanglement in the frequency domain. This method provides a light source capable of wide-range quantum sensing and quantum imaging or high-dimensional quantum processing.

## 1. Introduction

Entanglement is a prominent feature of quantumly indistinguishable states and a key resource for a variety of quantum devices and processes. An entangled state can be constructed from photon pairs by using photon degrees of freedom such as polarization, wave-vector, or frequency [1,2]. To realize the potential of quantum applications, the ways of generating entangled photon pairs need to be coherent and versatile. A typical method is based on spontaneous parametric down conversion (SPDC) [3], in which one pump photon is coherently split into a pair of correlated photons, a signal and an idler. A remarkable feature of SPDC light sources is that the polarization, momentum (wave-vector) and frequency are uniquely determined by energy and momentum conservation rules, so that one can have a quantum-correlated photon pair with high definition of polarization, momentum and/or frequency state.

One of the most popular schemes to generate the entangled state is to use the two symmetry non-collinear processes which create the polarization-entangled photon pairs with high momentum definition [1]. Though this is a simple scheme using a single pump and crystal, there is a limitation to using spatially separated photon pairs for quantum applications. For this reason, it is essential to develop a generation method of entangled photon pairs using the collinear SPDC process. In many cases, the collinear SPDC process has been realized with additional operations such as filtering post-selection [4], interferometry [2,5,6], and double-pass configuration [7,8]. These complicated schemes, however, typically have problems with the tunability of the output wavelength, the purity of the quantum entanglement [9,10] or compatibility with integrated applications [11]. A simple but promising scheme has been explored in many proof-of-principle experiments.

Recently, in order to overcome such technical challenges, it has been demonstrated that two nonlinear optical processes take place simultaneously in a cascaded-grating nonlinear crystal with two different poling periods, which can be used to generate frequency-entangled states with high polarization definition in the collinear geometry [12–15]. One significant point is that the photon pairs generated in the two-period crystal exhibit entanglement in the polarization and frequency domain by spatially separating the photon pairs via a polarization or frequency beam splitter. In particular, Kaneda *et al.* [15] demonstrated that two non-degenerate SPDC pairs produced in the two-period PPLN exhibit the non-degenerate HOM interference. They confirmed that the cascaded-grating nonlinear crystal can directly form the frequency-bin entangled photon pairs. On the other hand, the output wavelengths generated by these techniques are in a limited spectral range due to the phase-matching conditions. For the fabricated crystal, only a set of wavelengths can excite the non-degenerate symmetric SPDC process. To realize the full potential of these techniques, tuning of the SPDC wavelengths over a wide spectral range including the infrared region remains a pressing issue for quantum sensing and communication implementations.

Here, we report a novel way to generate entangled photon pairs in the frequency and polarization domains by using simultaneous collinear spontaneous parametric down conversion (s-SPDC) [16,17]. Periodically poled stoichiometric lithium tantalate (PPSLT) crystals have a relatively small difference in refractive index between extra-ordinary and ordinary light rays [18,19]. It allows two type-II phase-matching conditions to be satisfied with the identical pump frequency, crystal period, and temperature. Two symmetric non-degenerate type-II SPDC pairs are spectrally indistinguishable after separated by a polarization beam splitter [15]. Therefore, the s-SPDC pairs become directly entangled without any complicated procedure as mentioned above. Furthermore, a key feature of the s-SPDC method is that the signal and idler can be set to a wide range of arbitrary frequencies by tuning the pump frequency and crystal temperature. Thus, it can readily generate entangled photon pairs consisting of signals and idlers with wavelengths not only in the telecom region but also far apart in the visible and infrared regions. Our light source will bridge the gap between the demands and technical limitations on applicability for quantum sensing and communication.

## 2. Theory

In this section, we discuss the theory of simultaneous phase-matching pair generation using the special birefringence properties of PPSLT [18]. Suppose that two independent collinear type II quasi-phase matching conditions are simultaneously satisfied in the same crystal with one polling period. In this case, the phase matching condition and law of the energy conservation can be written as follows:

$$\frac{2\pi}{\Lambda} = \frac{2\pi n_o(\lambda_p, T)}{\lambda_p} - \frac{2\pi n_e(\lambda_s, T)}{\lambda_s} - \frac{2\pi n_o(\lambda_i, T)}{\lambda_i}, \quad (1a)$$

$$\frac{2\pi}{\Lambda} = \frac{2\pi n_o(\lambda_p, T)}{\lambda_p} - \frac{2\pi n_o(\lambda_s, T)}{\lambda_s} - \frac{2\pi n_e(\lambda_i, T)}{\lambda_i}, \quad (1b)$$

$$\frac{1}{\lambda_p} = \frac{1}{\lambda_s} + \frac{1}{\lambda_i}, \quad (1c)$$

where $\lambda_j$ is the wavelength of the pump ($j = p$), idlers ($j = i$), and signals ($j = s$) in vacuum, and $n_{e(o)}(\lambda_j, T)$ is the extra-ordinary (ordinary) refractive index of the crystal with wavelength $\lambda_j$ and temperature $T$. $2\pi n(\lambda_i, T)/\lambda_i$ denotes the wave vectors and $2\pi/\Lambda$ is the superlattice vector obtained from the periodicity $\Lambda$ for compensating the phase mismatch. Here, equations (1a) and (1b) are reversed in the polarization of the signal and idler waves. To find the set of $\lambda_p$, T, and $\Lambda$ that satisfy both equations, we subtracted equation (1a) from equation (1b) to derive the following conditional expression:

$$\frac{1}{\lambda_c} = \frac{n_e(\lambda_s, T) - n_o(\lambda_s, T)}{\lambda_s} = \frac{n_e(\lambda_i, T) - n_o(\lambda_i, T)}{\lambda_i}. \quad (2)$$

$1/\lambda_c$ is a condition parameter. In order to investigate Eqn. (2), we define a function $F_T(x) = \{n_e(x, T) - n_o(x, T)\}/x$. The existence condition for the solution of Eq. (2) is that for a given $\lambda_c$, $F_T(x) = 1/\lambda_c$ has multiple solutions in the transparent wavelength range 0.4-5 μm of the PPSLT crystal. This means that $F_T(x)$ has extreme values in this region. Figure 1(a) shows the curves of $y = F_T(x)$ with different temperatures in the range of 75-105 ℃. Here we used temperature dependent refractive indexes from Ref. [18] (Supplementary 1). The number of extreme points is highly dependent on the crystal temperature. We conclude that multiple solutions for $F_T(x) = 1/\lambda_c$ exist only for 78-103 ℃, where frequency-entangled photon pairs with high polarization definition can be generated. Note that the number of the solutions varies depending on the temperature and $1/\lambda_c$. The case of the temperature with more than two solutions are investigated in Supplementary 1.

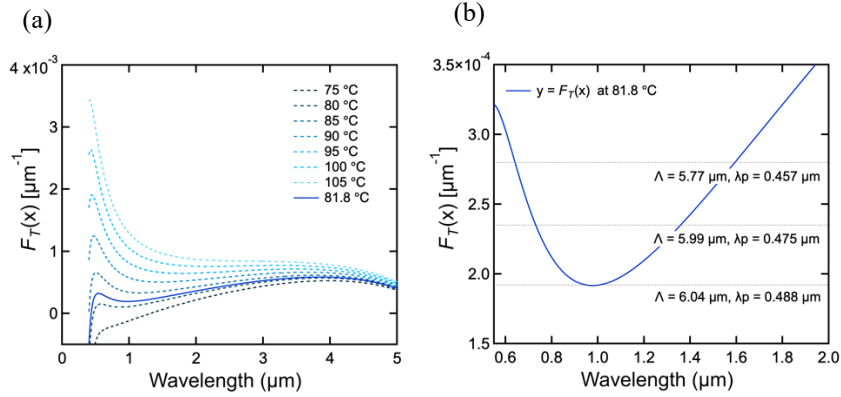

**Fig. 1.** (a) Calculated curves of $y = F_T(x)$ for PPSLT with different temperatures 75-105 ℃ based on the Sellmeier equations from Ref. [18]. The number and position of the solutions that satisfy Eqn. (2) vary as changing the temperatures. (b) Enlarged plot of the curve at 81.8 ℃. The output wavelengths and corresponding poling period and pump wavelength are arranged by changing the value of $1/\lambda_c$.

Now, we focus on the special case of 81.8 ℃ (solid curve in Fig. 1(a)) which corresponds to the experimental condition as described below. Figure 1(b) shows the enlarged plot around the minimum value of the solid curve. The dashed lines represent several $1/\lambda_c$s to determine the solutions. For example, for $1/\lambda_c = 1.919 \times 10^{-4}$, which is close to the minimum, the signal and idler are generated at 0.955 and 0.998 µm. This directly gives the pump wavelength $\lambda_p$ as 488 nm from Eqn. 1(c). The poling period Λ is obtained as 6.04 µm from Eqn. 1(a) or 1(b). As increasing $1/\lambda_c$, the output wavelengths are arranged to the widely-separated combinations as shown in Fig. 1(b). One essential feature is that the crystal temperature determines the tunability of the output wavelength. According to our calculation, the widest tunability can be obtained with the temperature at 84.7 ℃, in which the signal and idler wavelength can be arranged from the degenerate pair of 1.14 µm and 1.14 µm (Λ = 9.69 µm, $\lambda_p$ = 0.57 µm) to the non-degenerate pair of 0.55 µm and 3.78µm (Λ = 9.80 µm, $\lambda_p$ = 0.47 µm). It should be noticed that one could not set the crystal temperature, pump wavelength or poling period independently. We will discuss the case of the fixed crystal polling period in the next section and the case of fixed pump wavelength in the Supplementary 2.

Note that other nonlinear crystals such as periodically poled lithium niobate (PPLN) have no solution for Eqn. (2) due to the larger birefringence. The difference between $n_e$ and $n_o$ of PPLN crystals monotonically changes and is more than 0.05 in the transparent spectral region [20]. In such case, $F_T(x)$ also monotonically changes, which implies that there are no combinations of $\lambda_p$, $T$, and Λ to satisfy Eqn. (2). However, this is not the case for some crystals such as PPSLT crystal. The difference between $n_e$ and $n_o$ of PPSLT crystals is not only less than 0.004 but also does not monotonically change [18]. These features allow $F_T(x)$ to have several extremum points in the visible and infrared region, which means that there is a set of

$\lambda_p$, $T$, and $\Lambda$ that satisfies both equations above. The detail calculations for PPLN crystals are shown in Supplementary 1.

The output quantum state $|\varphi\rangle$ of the simultaneous SPDC pairs can be described as follows:

$$|\varphi\rangle = \frac{1}{\sqrt{2}}\left(|H,\omega_1\rangle|V,\omega_2\rangle + e^{i\theta}|V,\omega_1\rangle|H,\omega_2\rangle\right). \qquad (3)$$

Here, the quantum states are described using horizontal (H) or vertical (V) polarization modes, and the angular frequency $\omega_{1(2)}$ in place of the wavelength $\lambda_{s(i)}$. $\theta$ is the relative phase between two quantum states, which is derived from the temporal gap caused in the generation process between signals and idlers and is arranged by the external delay line after separating the photons by a polarization beam splitter (PBS) or dichroic mirror (DM). By using PBS or DM, $|\varphi\rangle$ is spatially separated and reduced to a frequency-bin or polarization-bin state [15]. In our experiment, we separated the photon pairs by using a PBS and obtained a frequency-bin state as follows:

$$|\varphi_F\rangle = \frac{1}{\sqrt{2}}\left(|\omega_1\rangle_H|\omega_2\rangle_V + e^{i\theta}|\omega_2\rangle_H|\omega_1\rangle_V\right) = \frac{1}{\sqrt{2}}\left(|\omega_1\omega_2\rangle_{HV} + e^{i\theta}|\omega_2\omega_1\rangle_{HV}\right). \qquad (4)$$

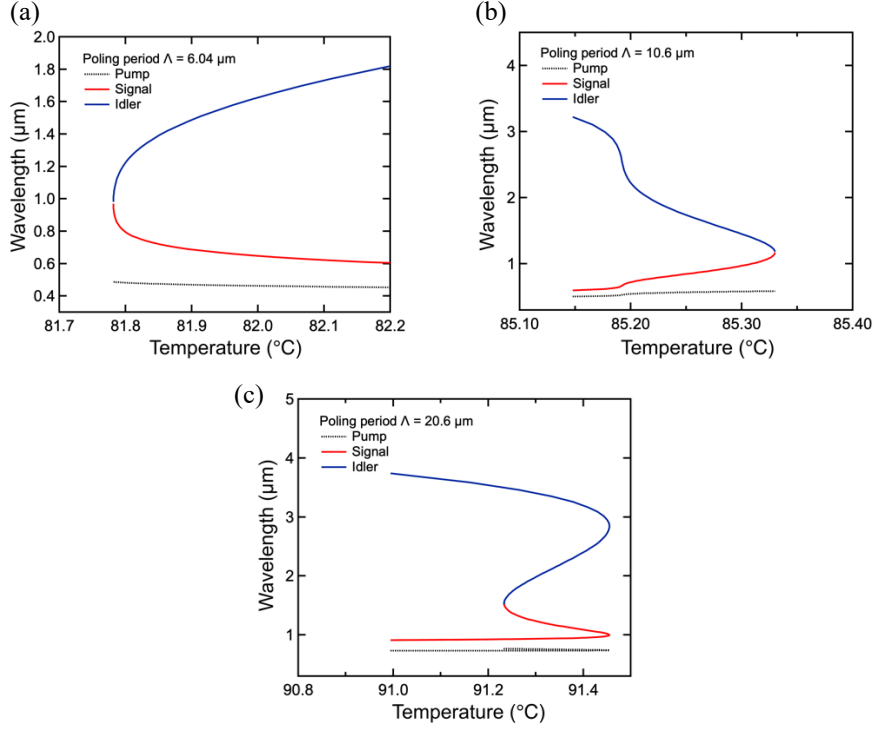

**Fig. 2.** Phase matching conditions of the type-II s-SPDC processes in PPSLT calculated using the Sellmeier equations from Ref. [18]. Output wavelengths (signal: red line, idler: blue line) and pump wavelengths (dotted line) are shown as function of the crystal temperature for several polling periods: (**a**) $\Lambda$=6.04 µm, (**b**) $\Lambda$=10.6 µm, (**c**) $\Lambda$=20.6 µm.

## 3. Wavelength tuning

To investigate the potential tunability of the type-II s-SPDC processes, we calculated the possible combinations of the pump wavelengths and crystal temperatures while using the fixed poling period. Figure 2(a)-(c) show the output wavelengths and pump wavelengths as function of the crystal temperature for several polling periods. The red, blue and black solid curves indicate the signal, idler and pump wavelengths, respectively. As shown in Fig. 2(a), the s-SPDC process with the polling period of $\Lambda = 6.04$ μm has a potential coverage in the visible and near infrared region. Especially, the degenerate pairs are generated around 1 μm with the temperature at 81.8 °C. In the case of $\Lambda = 10.6$ μm, the idler wavelengths can be arranged even into the infrared region. Moreover, by setting $\Lambda = 20.6$ μm, the degenerate pairs are generated in the telecom band around 1.5 μm, and the non-degenerate pairs can be widely separated in the visible and infrared region. It should be noticed that the tuning range of the pump wavelength and crystal temperature is quite narrow, which means that the pump wavelength and the crystal temperature should be finely tuned for generating a certain combination of signal and idler. Additionally, this tuning capability can be also achieved by changing the poling period and crystal temperatures while using the fixed pump wavelengths. In this case, the tunability of the output wavelength is realized by setting the temperature to satisfy Eqn. (1a) and (1b) with the poling period in a certain range. The calculated results of these tuning arrangement are shown in Supplementary 2.

To experimentally demonstrate the tunability discussed above, we measured the s-SPDC signal spectra in the case of operating at a fixed pump wavelength at 488 nm with varying crystal temperatures and poling periods. Figure 3(a) shows the experimental setup to measure signal spectra using a wire grid polarizer (WGP) for analyzing the distribution of the polarization modes. A PPSLT crystal was excited by a 488 nm continuous wave laser diode (linewidth: 1 nm) with an average power of 1 mW. We employed two 11-mm-long fan-out PPSLT crystals (Oxide Corporation) for setting a variety of the poling periods. The output signals were collected and introduced into a silicon CCD spectrometer after collimated by a 50-mm-focal lens. We used appropriate frequency filters, depending on the situation. The WGP extracted a certain linear polarization mode.

Figure 3(b) shows the measured spectra in H- and V-polarization (blue and green curve, respectively) for two conditions $\alpha$ and $\beta$. The peaks were observed in both polarizations, which is consistent with the theoretical prediction of s-SPDC processes. All spectrum shows relatively wide spectral bandwidth due to a broadening effect mainly caused by a finite 1 nm spectral linewidth of the pump light. In the case of $\alpha$, we used the pooling period of $\Lambda = 6.04$ μm, which is the near degenerate condition where the frequencies of the signal and idler waves are close to each other, we originally observe only one peak at 75 °C. For the quantum interference experiment of the frequency bin mode, we need to separate the signal and idler waves, which have explicitly narrower linewidth than the wavelength separation. Therefore, we employed three spectral filters to be set after the PPSLT in order to extract only the corresponding pair of signal and idler pairs from the broad s-SPDC emissions [21]. As shown in Fig. 3(b), we picked up a pair of signals (0.955 μm) and idlers (0.998 μm). The details of the broadening effect and filtering manipulation are discussed in Supplementary 3. In the case of $\beta$, we used the poling period of $\Lambda = 7.83$ μm and observed the signal wave at 0.637 μm at the crystal temperature of 76 °C. It should be mentioned that in both cases of $\alpha$ and $\beta$ the intensity of the H-polarization is stronger than that of the V- polarization. This is because the type-0 process (all photon is in the H-polarization) is also spontaneously caused in the PPSLT and contributes to the counts of the H-polarization. The discussion of the polarization analysis for the signals is shown in Supplementary 4.

In order to confirm the validity of the theory, we simulated the s-SPDC spectra using Eqn. (4) in Ref. [16] with Eqn. (1a). Figure 3(c) shows the calculated spectra obtained with the same pump wavelength and crystal poling periods. The crystal length was taken as the same value used in the experiments. The phase matched temperatures are automatically determined in both cases as mentioned in the previous section. However, the temperatures are found to shift systematically by 7-8 K from the experimental values. Possible causes of this shift include the crystal being strictly different from the one that determined the Sellmeier equation [18], local heating due to the pump laser light, and temperature gradients, but it has not been clarified yet which of these factors is responsible. Except for this temperature shift problem, the center wavelengths of calculated signal spectra have a good agreement with measured spectra shown in Fig. 3(b). One remarkable point is that the corresponding idles in $\beta$ exists at 2.08 μm in the mid-infrared region. In other word, by arranging the crystal temperatures and poling periods with the fixed pump wavelength 488 nm, one can cover wide spectral range even in the infrared region. Thus, the type-II s-SPDC generator can be served as a widely tunable frequency-entangled photon source covering from the visible to the infrared region.

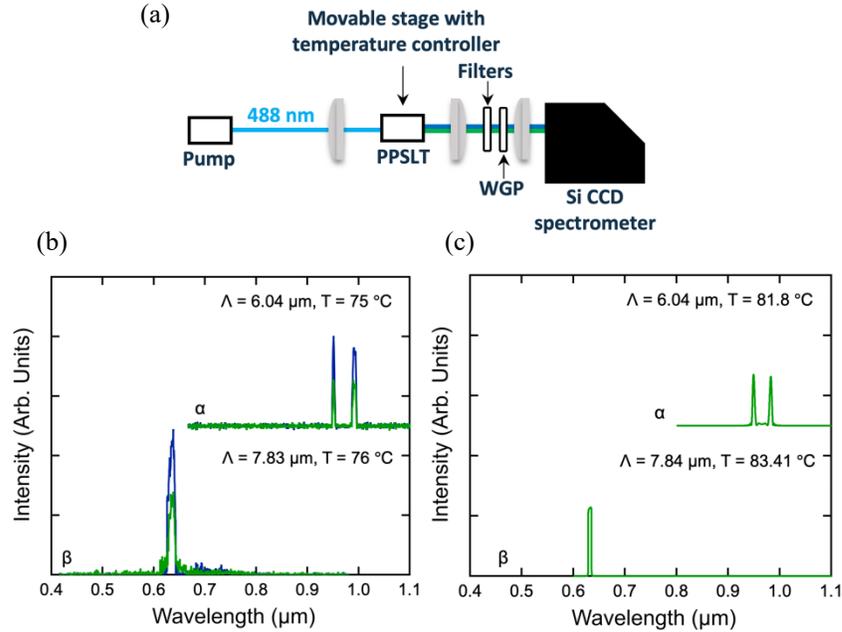

Fig. 3. (a) Experimental setup for measuring the SPDC signal spectra. PPSLT was mounted on the movable stage. Wire grid polarizer (WGP) was inserted for extracting a certain linear polarization mode before coupled into the spectrometer. (b) Measured spectra with different crystal periods and temperatures in the horizontal (blue) and vertical (green) polarization mode. (c) Theoretical simulations of s-SPDC spectra for pump wavelength of 488 nm using Eqn. (1a) based on Eqn. (4) in Ref. [16] and Sellmeier equations from Ref. [18].

## 4. Hong-Ou-Mandel interference

To demonstrate entanglement between two SPDC pairs, we constructed a HOM interferometer in the frequency-bin configuration for the near degenerate pairs ($\alpha$ in Fig. 3(b)). The experimental layout of the HOM interference measurement is shown in Fig. 4. We used the same pump wavelength and crystal parameters as is for $\alpha$ in Fig. 3(b). After the residual pump was removed by 700-nm-long-pass filters (LP700s), the SPDC pairs from the PPSLT were introduced into a Michelson interferometer with a polarization beam splitter (PBS1) and quarter wave plates (QWPs) [15]. PBS1 spatially separated signals and idlers into two blanches, respectively, and the QWP was served for SPDC pairs to pass into another PBS1 port by rotating the polarization after reflected by the mirror. One of the two blanches in the interferometer has a delay line for arranging the relative phase between two SPDC pairs by changing the optical length. After the interferometer, the polarizations of two pairs were changed to a diagonal mode (45 degrees) by a half wave plate (HWP) set at 22.5 degrees in order to equally divide both pairs into two output ports of PBS2, which functioned as a non-degenerate HOM interferometer. The photons were focused into single-mode-fiber coupling (SMF) systems and detected by silicon-based single photon-counting avalanche diodes (Si-SPADs). Two Si-SPADs were connected to a time-correlated single photon counter (TCSPC) module to measure coincidental events in the two PBS2 branches. The integration time was 50 seconds and the temporal window of the entire coincidence measurement system was set at 4 ns. The typical count rate of each port was around $3500 \pm 59$ cps after the filtering system to remove the excess spectral broadening and extracted only discrete-color photon pairs as described in Supplementary 3. For measuring the HOM interference, the coincidence events $C_O$ between two detecting ports were recorded in the TCSPC as changing the delay line with step size of 50 nm. To convert the coincidental counts into normalized counts, we also measured the coincidence counts $C_F$ at the position of the delay far from the temporal origin, where no interference occurred. Note that we measured the coincidence counts with the center wavelengths of two peaks that does not satisfy energy conservation with the pump, in which we found there were little coincidence counts. It follows that the filtering system was properly placed for extracting the narrow s-SPDC components. We also estimated the conversion efficiency was $1.4 \times 10^{-12}$ using two single counts and coincidence counts [22], which was sufficient for the entangled photon-pair generation applications even though it was one order of magnitude smaller than to the generation schemes using other nonlinear crystals such as PPLNs [15].

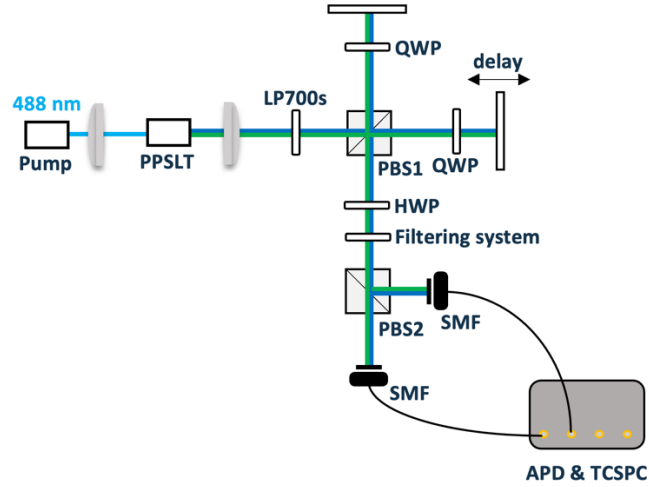

**Fig. 4.** Schematic layout of the experiment. LP700s: long pass filters. PBS: polarization beam splitter. QWP: quarter wave plate. HWP: half wave plate. SMF: single mode fiber coupling. APD: avalanche photo diode. TCSPC: time correlated single photon counter.

Figure 5(a) shows the measured HOM interferogram of the normalized coincidental events $C_N = C_o/2C_F$. In Fig. 5(a), the counts oscillated in a certain range, with a frequency corresponding to the spectral difference (13.5 THz) between the two peaks as shown in Fig. 3(b). This interferogram is theoretically described by the non-degenerate interference equation for spatial beating [2] as follows:

$$C_N = \frac{1}{2}\left(1 - V\cos(\delta\omega t + \theta)\left(1 - \left|\frac{t}{t_c}\right|\right)\right) \quad \text{for } t < t_c. \quad (5)$$

Here, $V$ is interference visibility and $\delta\omega$ is beating frequency determined by the wavelength difference between the signal and idler. $t_c$ is coherence time determined by the spectral width of the SPDC signals and idlers. By fitting the fringe to Eqn. (5), we estimated that $V$, $\delta\omega$, $\theta$, and $t_c$ were $0.82 \pm 0.01$, $13.5 \pm 0.1$ THz, $0.205 \pm 0.001$ rad, and $0.69 \pm 0.01$ ps, respectively. The estimated value of $\delta\omega$ is in a good agreement with the one predicted from the measured spectral difference. The value of $t_c$ matched the bandwidth of the signals. The visibility indicates indistinguishability of the s-SPDC pairs, which is limited by a broad phase-matching window due to the spectral width of the SPDC pump, and excess photons from the type-0 SPDC process. The spectral filtering manipulation used in the measurement decreased the total throughput of the system, and the type-0 SPDC pairs increased the accidental coincidence counts, which limited the visibility. Moreover, we estimated the density matrix $\rho_F$ of the s-SPDC pairs in the frequency domain. According to the method used in Ref. [2], the matrix is constructed by parametrization as follows:

$$\rho_F = p|\omega_1\omega_2\rangle\langle\omega_1\omega_2|_{HV} + (1-p)|\omega_2\omega_1\rangle\langle\omega_2\omega_1|_{HV}$$
$$+ \frac{V}{2}\left(e^{i\theta}|\omega_1\omega_2\rangle\langle\omega_1\omega_2|_{HV} + e^{-i\theta}|\omega_1\omega_2\rangle\langle\omega_1\omega_2|_{HV}\right). \tag{6}$$

Here, the balance parameter $p$ is the probability of $|\omega_1\omega_2\rangle_{HV}$, which is obtained from the ratio of the coincidence events of $|\omega_1\omega_2\rangle_{HV}$ and $|\omega_2\omega_1\rangle_{HV}$. In our setup, we estimated $p = 0.538 \pm 0.002$ by comparing the number of single counts in each port.

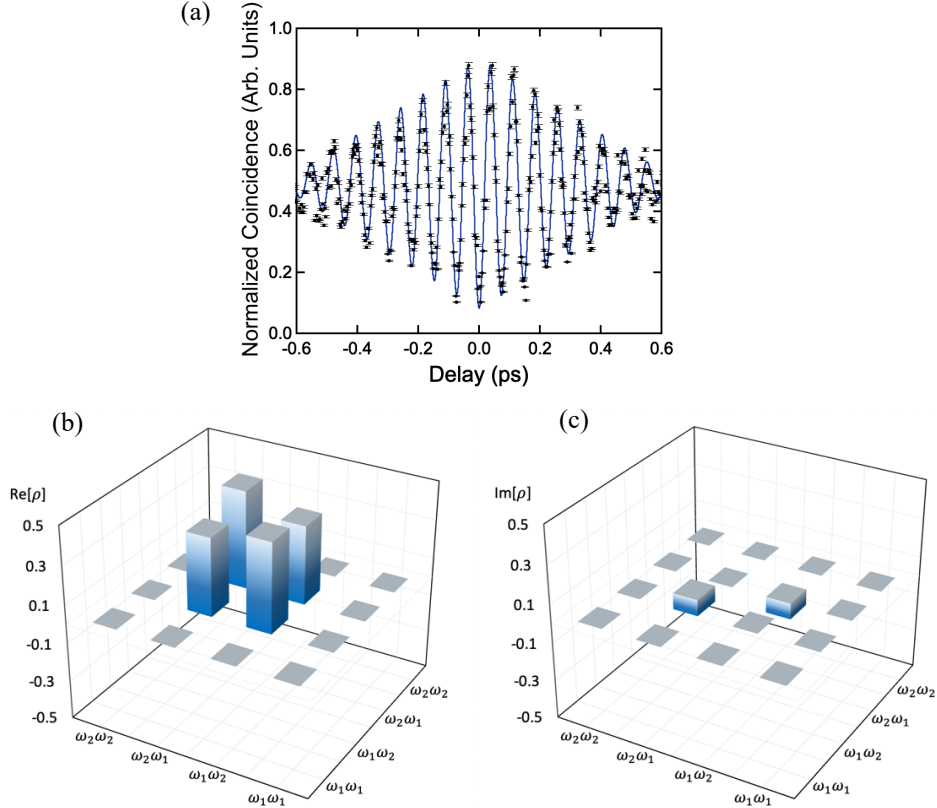

**Fig. 5.** (**a**) Measured Hong-Ou-Mandel (HOM) interferogram for type-II s-SPDC frequency-entangled photon pairs in PPSLT (pump wavelength: 488 nm, crystal temperature: 75 ℃, polling period: $\Lambda = 6.04$ μm). The estimated beating frequency and visibility of the fringe are 13.5 THz and 0.82. (**b**) (**c**) Real and imaginary parts of the reconstructed density matrix elements in the frequency domain.

Figures 5(b) and 5(c) show real and imaginary parts of the reconstructed density matrix elements using the parameter $V$, $\theta$, and $p$ estimated by the fitting curve shown in Fig. 5(a). As Eqn. (6) indicates, the non-diagonal components in the density matrix correspond to the purity and relative phase of entanglement between photon pairs. In Fig. 5(b) and 5(c), we obtained uneven but explicit non-diagonal components in the real part, and a finite but small value of non-diagonal components in the imaginary part. Theoretically, ideal entangled states have $V/2 = p = 0.5$, which means that the deviation from the ideal states may come from the fact that $V/2$ and $p$ is not 0.5. The fidelity to the ideal entangled states was $F = 0.90 \pm 0.001$, even more than the classical limit 0.5 [23]. From these results, we conclude that our s-SPDC process successfully generated entangled photon pairs. Note that the visibility $V$ is maximized at $p = 0.5$ for $V/2 \leq \sqrt{p(1-p)}$. Therefore, the indistinguishability can be improved by making $p$ come close to 0.5 if we employed adequate optical components such as the PBS perfectly splitting the polarization modes over the broadband range. Moreover, the filters we used also decreased the visibility because the extracted spectral coverage cannot be perfectly proper. Using a narrow pump would improve indistinguishability.

## 5. Conclusion

We proposed a way to generate widely-tunable frequency entangled photon pairs with high polarization definition in a single-pass configuration by using the simultaneous phase-matching conditions of a non-degenerate type-II SPDC. In this technique, the specific birefringence in the nonlinear medium is the key that allows the simultaneous processes. We investigated the capability of a PPSLT crystal in generating entangled photon pairs over a wide range of frequencies including the infrared domain. One significant aspect of the method is that the pump can be of any wavelength, which allows the entangled photon pairs to be produced with arbitrary spectral coverage, e.g., on telecom wavelengths or within the molecular fingerprint range.

Moreover, we experimentally demonstrated entanglement between two SPDC pairs in the frequency domain. Our experimental setup used a filtering system for extracting narrow bandwidths, which lowered the purity of the entangled state, though we obtained a clear spatial beating of the coincidental events. Using a narrower-band pump laser source, long crystal or more accurate beam splitter would allow us to have more discrete spectrum and increase the purity of entanglement. Additionally, multi-wavelengths filtering or multi-grating crystal can readily multiply the entanglement dimensions by increasing the frequency modes [24,25]. Therefore, our method may propel the development of not only multi-dimensional quantum protocols using entangled states for computing and communication, but also techniques of quantum sensing in a wide spectral region.

**Acknowledgments.**

The Authors gratefully thank Dr. Kento Uchida, Prof. Dr. F. Kaneda, and Prof. Dr. K. Edamatsu for fruitful discussions. This work is supported by the Quantum Leap Flagship Program (MEXT Q-LEAP) (JPMXS0118067634) of the Ministry of Education, Culture, Sports, Science and Technology and the Public/Private R&D Investment Strategic Expansion Program (PRISM) of the Cabinet Office, Government of Japan. This work was also supported by the WISE Program of the Ministry of Education, Culture, Sports, Science and Technology, and by JSPS KAKENHI Grant Numbers 21J23205.

**Disclosures.**

The authors declare no conflicts of interest.

**Data availability.**

Data underlying the results presented in this paper are available in Dataset 1, Ref. [26].

**Supplemental document.**

See the Supplement Document for supporting content.

**Figure Captions**

Figure 1

(**a**) Calculated curves of $y = F_T(x)$ for PPSLT with different temperatures 75-105 ℃ based on the Sellmeier equations from Ref. [18]. The number and position of the solutions that satisfy Eqn. (2) vary as changing the temperatures. (**b**) Enlarged plot of the curve at 81.8 ℃. The output wavelengths and corresponding poling period and pump wavelength are arranged by changing the value of $1/\lambda_c$.

Figure 2

Phase matching conditions of the type-II s-SPDC processes in PPSLT calculated using the Sellmeier equations from Ref. [18]. Output wavelengths (signal: red line, idler: blue line) and pump wavelengths (dotted line) are shown as function of the crystal temperature for several polling periods: (**a**) Λ=6.04 μm, (**b**) Λ=10.6 μm, (**c**) Λ=20.6 μm.

Figure 3

(**a**) Experimental setup for measuring the SPDC signal spectra. PPSLT was mounted on the movable stage. Wire grid polarizer (WGP) was inserted for extracting a certain linear polarization mode before coupled into the spectrometer. (**b**) Measured spectra with different crystal periods and temperatures in the horizontal (blue) and vertical (green) polarization mode. (**c**) Theoretical simulations of s-SPDC spectra for pump wavelength of 488 nm using Eqn. (1a) based on Eqn. (4) in Ref. [16] and Sellmeier equations from Ref. [18].

Figure 4

Schematic layout of the experiment. LP700s: long pass filters. PBS: polarization beam splitter. QWP: quarter wave plate. HWP: half wave plate. SMF: single mode fiber coupling. APD: avalanche photo diode. TCSPC: time correlated single photon counter.

Figure 5

(**a**) Measured Hong-Ou-Mandel (HOM) interferogram for type-II s-SPDC frequency-entangled photon pairs in PPSLT (pump wavelength: 488 nm, crystal temperature: 75 ℃, polling period: Λ = 6.04 μm). The estimated beating frequency and visibility of the fringe are 13.5 THz and 0.82. (**b**) (**c**) Real and imaginary parts of the reconstructed density matrix elements in the frequency domain.